# Multipath CNN with alpha matte inference for knee tissue segmentation from MRI

Sheheryar Khan, *Member, IEEE*, Basim Azam, Yongcheng Yao, Weitian Chen

*Abstract* — Precise segmentation of knee tissues from magnetic resonance imaging (MRI) is critical in quantitative imaging and diagnosis. Convolutional neural networks (CNNs), which are state of the art, have limitations owing to the lack of image-specific adaptation, such as low tissue contrasts and structural inhomogeneities, thereby leading to incomplete segmentation results. This paper presents a deep learning–based automatic segmentation framework for knee tissue segmentation. A novel multipath CNN-based method is proposed, which consists of an encoder–decoder-based segmentation network in combination with a low rank tensor-reconstructed segmentation network. Low rank reconstruction in MRI tensor sub-blocks is introduced to exploit the structural and morphological variations in knee tissues. To further improve the segmentation from CNNs, trimap generation, which effectively utilizes superimposed regions, is proposed for defining high, medium and low confidence regions from the multipath CNNs. The secondary path with low rank reconstructed input mitigates the conditions in which the primary segmentation network can potentially fail and overlook the boundary regions. The outcome of the segmentation is solved as an alpha matting problem by blending the trimap with the source input. Experiments on Osteoarthritis Initiative (OAI) datasets and a self-prepared scan validate the effectiveness of the proposed method. We specifically demonstrate the application of the proposed method in a cartilage segmentation–based thickness map for diagnosis purposes.

## I. INTRODUCTION

Knee osteoarthritis (OA) is a common chronic and degenerative musculoskeletal disease that affects a significant population worldwide [1]. Clinically, knee OA is characterized by structural changes and degeneration in knee joint tissues, specifically in articular cartilages around bones [2]. Magnetic resonance imaging (MRI) is the preferred method for generating desired tissue contrast and offers multiple imaging options through a variety of acquisition protocols [3][4][5]. However, acquired MR images often exhibit qualitative diversity among examinations that vary in imaging protocols, subjects, and hardware.

Knee OA is generally considered to be an inflammatory and biomechanical whole-organ disease in which several factors, such as joint shape and tissue degeneration, have a strong influence on disease progression [2]. Non-invasive characterization of these features by MRI is valuable for diagnosis, prognosis, and treatment planning of knee OA [5]. Determination of compositional and morphological changes in articular cartilage and neighboring tissues is of foremost interest in evaluating therapeutic procedures and demonstrating the pathogenesis of knee OA [6]. To carry out these investigations from MRI, segmentation of tissues is often required. However, manual delineation of knee tissues is time-consuming and can be impractical when processing large cohorts in routine clinical applications. Therefore, automatic and robust segmentation of knee tissues is needed.

Machine learning and deep learning (DL) models have laid a solid foundation for modeling complex problems in knee MR image computing and have provided several exciting avenues of research to understand clinical conditions [7][8][12][13]. Comprehensive reviews of the latest advances in automatic segmentation methods can be found in [3][9]; in particular, convolutional neural networks (CNNs) have shown promise in their performance in knee tissue segmentation from MRI scans. Among CNNs, Seg-Net [10] and U-Net [11] have gained considerable popularity in deep CNN models owing to their effectiveness in automatic segmentation of cartilage and adjacent tissues in MRI [12][13]. In U-Net-based architectures, the features from earlier encode layers are mapped onto the later layers, resulting in an enhancement of the learning capability in semantic segmentation. Therefore, U-Net-based architectures are widely accepted in knee tissue segmentation. Several improved versions have been proposed; for example, 2D U-Net with skip connections for knee cartilage segmentation was presented in [12], whereas Seg-Net with postprocessing was proposed [13] for a similar problem—this approach incorporates 3D simplex deformable modeling for bone and cartilage segmentation in knee MRI and produces smooth deformation along bone and cartilage boundaries. Meanwhile, an approach that combined a CNN-based conditional random field model with spatial proximity was presented by Zhou et al. in [14], to improve the precision of the predicted labels. All the reported approaches based on DL methods significantly

This work is supported by a grant from the Innovation and Technology Commission of the Hong Kong SAR (Project MRP/001/18X), and Faculty Innovation Award from the Faculty of Medicine, the Chinese University of Hong Kong.
Sheheryar Khan, Yongcheng Yao and Weitian Chen are with the Department of Imaging and Interventional Radiology, CUHK lab of AI in radiology (CLAIR), Chinese University of Hong Kong, Shatin N.T., Hong Kong. (e-mail: sheheryar1984@gmail.com; yongcheng.yao@cuhk.edu.hk; wtchen@cuhk.edu.hk); (Corresponding author: Weitian Chen)
Basim Azam is with Centre for Intelligent Systems, Central Queensland University, Brisbane, Australia. (e-mail: b.azam@cqu.edu.au)

improved the segmentation when compared with the conventional machine learning approaches; however, modeling the patient-specific structural variations in cartilage and meniscus tissues is challenging. The lack of labeled data to cover the small local shape variations in affected OA patients is a contributory factor that prevents the generation of precise tissue regions during inference.

We approach the problem of knee tissue segmentation by using a procedure that leverages a secondary input from data and is based on low rank reconstruction of the source input. Our method is aimed at addressing segmentation challenges, particularly those arising from inconsistencies in anatomical regions, differences in signal intensities, partial voluming and lack of tissue boundaries. As three-dimensional (3D) MR images of adjacent slices have a strong correlation between adjacent slices, effective utilization of this information is a promising way to model the intrinsic structural similarities of the tissues in the slice dimension. Intuitively, a block-wise low rank (LR) reconstruction of 3D MRI has an overall effect of eliminating information redundancy and highlighting the tissue regions. In terms of segmentation, in particular, CNNs trained on LR input can capture this information and can prevent the missing boundaries issue. Inspired by the LR reconstruction of knee MRI to highlight tissue inconsistencies, we design a multipath CNN strategy to segment the semantic information of knee tissues from MRI. As knee tissues are composed of thin boundaries in a single slice and are connected in the slice dimension, any under segmentation from LR reconstruction can influence the outcome of segmentation; therefore, fusing the outcomes of both CNNs can highlight the confident and proximate regions. Therefore, we propose *trimap* generation as a mechanism for the fusion of the outcomes of the CNNs. The procedure is shown in Fig. 1; both paths contribute to generating the intermediate segmentation result in the form of a *trimap* that consists of segmentation regions corresponding to high, low and medium confidence. Boundary regions and edges are overlooked at the initial stage of segmentation (c), whereas LR-based segmentation is used to capture the proximity regions and generate a *trimap* (d). The resulting labels from each path can later be fused into image matting along with actual input inference for decision making in regions of uncertainty (f). The additional learned feature space with an effective matting-based outcome provides improved segmentation performance and acceptable generalization ability.

### A. Literature Review and Context

Knee tissue segmentation of MR images has been long studied. A variety of semi-automatic and automatic approaches exist, with an increased emphasis on automatic generalizable methods in recent times [12]. Statistical shape models (SSM) gained considerable popularity in cartilage segmentation; they leveraged anatomical knowledge in terms of geometric priors [16]. The technique performs well even for low-contrast images. However, this method involves the use of a heuristically designed appearance model that fits the MRI data. A major drawback of the approach is the poor generalization to image domain variations. To overcome this issue, random forest regression voting [18] was introduced in 2D and 3D forms to model the appearance information from the input.

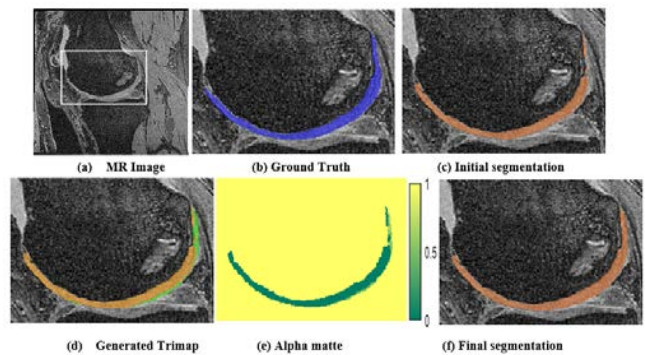

Figure 1: Demonstration of the proposed segmentation method. From left to right (input image with ROI marked cartilage region): ground truth region overlaid on cropped cartilage surface (blue), segmentation region predicted by the baseline network. Bottom row shows the regions marked by trimap generation through LR network, alpha matte of the cartilage with probability map, and final segmentation outcome from alpha matting.

DL-based CNN methods have become popular in medical imaging in recent years. However, the field still lacks efficient and reliable methods that can model the patient-specific variability in tissues in musculoskeletal MRI. Recent progress in CNN-based segmentation of knee cartilage and meniscus shows that it can outperform conventional atlas-based methods [9]. Segmentation of tibial cartilage from MRI was presented in [19]; 2D CNNs were trained on the axial, coronal and sagittal planes, and the final segmentation was produced by combining the outcomes from each orthogonal plane. The method outperforms handcrafted feature-based methods and single-plane CNNs. An enhanced version of this approach is presented in [13]; 3D simplex deformable modeling for smooth deformation of cartilage and bone segmentation is used to obtain 3D masks. In this case, the extracted image information is bound to the localized region as CNNs lack the context related to adjacent voxels. Integrating 2D and 3D information in CNN architecture was shown to be successful in meniscus segmentation [21]. The authors proposed the integration of 2D CNNs and 3D SSMs with 3D CNNs such that the structural information can be preserved before segmentation with 3D CNNs. However, the primary focus of this approach was only on meniscus tissues.

Inspired by the structural consistency of SSMs, Ambellan et al. [17] embedded SSMs as a postprocessing step in combination with 2D and 3D CNNs. An SSM performs well for bone segmentation and improves the outcome by filling the holes from the previous step. A similar approach that involved the addition of a conditional random field (CRF) was proposed in [20]. This method combines the CNNs with a CRF to achieve improved postprocessing.

Despite all these efforts, the main objective of the previous approaches was to enhance the capabilities of the network by using suitable postprocessing and correcting strategies, which are limited to the specific data and suit the predefined bias of the training data; there was little focus on learning these spatial inhomogeneities directly from data. It is likely that above approaches would be less effective in case the knee tissue pathologies turn out to be different.

### B. Contributions

Based on the discussion above, the spatial consistency of knee tissues is hard to capture. In this regard, our main contributions are the following:

1) We propose to address the knee tissue segmentation from MRI in a multipath CNN framework, in which features from both the source and their low rank reconstruction images are extracted.

2) We define a block-wise tensor decomposition and reconstruction to effectively model the structural non-homogeneities of the 3D MRI tensor and demonstrate the importance of learning the LR path through model design.

3) We formulate automatic *trimap* generation with uncertainty at low-confidence boundary regions and solve for alpha matte for the final prediction of labels while incorporating the source input again for inference.

We present an evaluation of the proposed method and examine the performance in relation to the key anatomical locations that are clinically relevant. We also explore the potential application of the proposed method by examining the segmentation and thickness map for a self-collected OA patient scan.

To the best of our knowledge, this is the first attempt to model the semantic segmentation for knee MRI with multipath CNNs and image matting in a fully automatic manner.

The remainder of this paper is organized as follows. In Section II, we describe the material and methods utilized in our proposed framework, with a complete outline of our segmentation approach. In Section III, we describe the experiments performed and present the results of the proposed segmentation algorithm. In Section IV, we discuss additional applications of the proposed method. Finally, we present the conclusion in Section V.

## II. METHODOLOGY

From the CNN-based knee MRI segmentation approaches, models that preserve shape and structure are found to be effective; they retain the best available information from input sequences. We propose a novel pipeline for knee MRI segmentation, specifically for tissues such as cartilage and meniscus, in terms of the geometric features.

### A. Outline of Approach

The proposed framework consists of three major components, as shown in Fig. 2. The input module (a) comprises source MRI scans and their block-wise low rank reconstruction, which is achieved through the Tucker decomposition of the consecutive stacked MRI slices in the slice dimension. The second stage (b) utilizes these distinct inputs and trains the U-Net-based CNN architectures to learn the comprehensive features. The first path, which feeds in the source MRIs, produces the segmentation maps that directly correspond to the spatial location of tissues, whereas the second path feeds in the LR reconstructed MRIs and generates slightly different segmentation maps due to low rank characteristics, such that non-static structures in the slice dimension are even more prominent and can be easily segmented by this path. In contrast to conventional voting schemes, we introduce the label fusion mechanism through image matting. As shown in module (c), we address the segmentation-map fusion problem and propose a *trimap* generation scheme that helps determine the most confident regions from both segmentation maps. In this scheme, the actual input MR image is considered again in the image matting process to evaluate whether the *trimap* regions belong to the desired tissue region or to the background region. The proposed framework constructs a segmentation label that supports the features from the true spatial regions and the features derived from the slice dimension simultaneously; therefore, when the targeted tissue is challenging to segment in a certain slice and results in some regions being missed, the corresponding path helps fill that information through low rank reconstruction. This helps provide an extra clue, which is modeled through *trimap* generation as an equally probable region which is finally addressed through the matting solution.

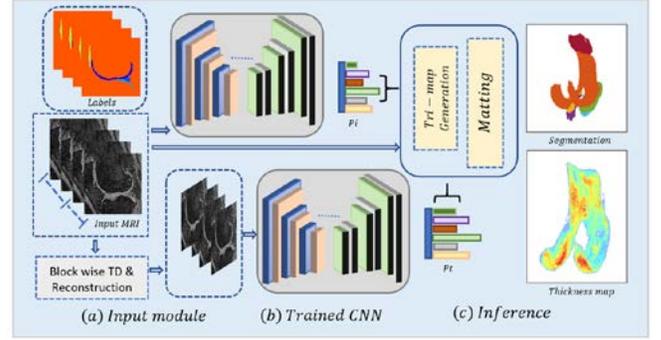

Fig. 2. The block diagram illustrating the framework of the proposed fully automatic segmentation method.

### B. Tucker Decomposition

Tensor decomposition through Tucker decomposition provides a powerful low rank approximation approach to decompose an input image representation into a set of projection matrices and a core tensor. Tucker decomposition has many applications in computer vision tasks, such as multispectral image restoration [24] and MRI T1 mapping [25]. Inspired by the success of tensor decomposition approaches in various fields and their low rank properties, we introduce a block-based Tucker decomposition and reconstruction stage to describe interdimensional tissue structures that can be learned in a parallel CNN path. The flowchart of the stage is shown in Fig. 3. A 3D MR image volume is subdivided into blocks of equal sizes, with consecutive MRI slices. The MRI block forming a third-order tensor is decomposed into a core tensor and mode matrices, which is then reconstructed by truncating the rank in the slice dimension.

For an $N$-order tensor $X \in \mathbb{R}^{I_1 \times I_2 \times \cdots \times I_N}$, Tucker decomposition produces $N$ orthogonal bases $\{U^{(n)} \in \mathbb{R}^{I_1 \times I_2 \times \cdots \times I_N}\}_{n=1}^{N}$ and a core tensor $Y \in \mathbb{R}^{R_1 \times R_2 \times \cdots \times R_N}$, where the rank associated with $X$ is $R_n \leq I_N$. Both the core tensor $Y$ and the input tensor can be formulated from the orthogonal matrices as:

$$Y = X \times_1 U^{(1)^T} \times_2 U^{(2)^T} \times_3 \ldots \times_N U^{(N)^T};$$

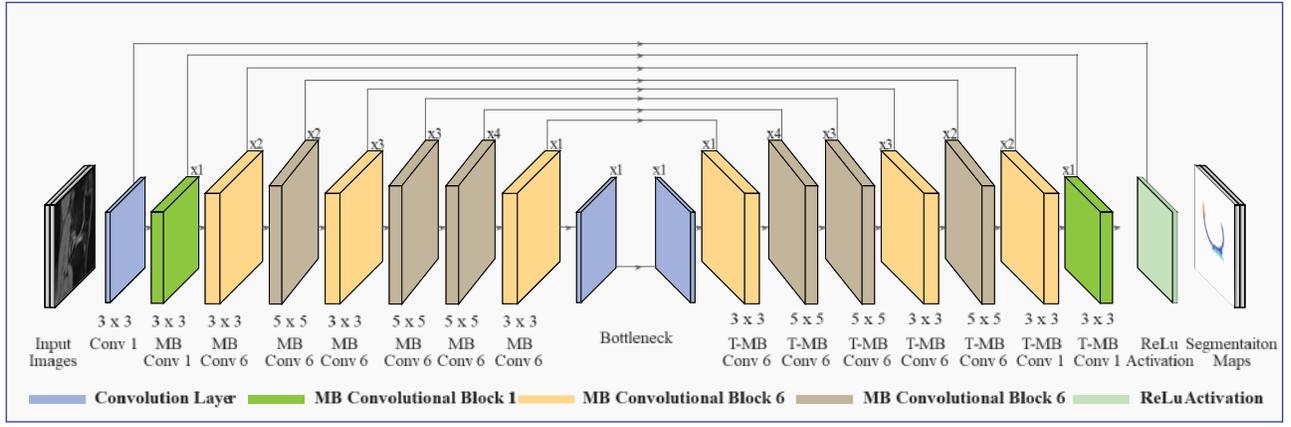

Figure 4: Block diagram of the CNN architecture used in the proposed framework. At each stage of the encoder and decoder, the input image is convolved progressively and rescaled by a factor of 2.

$$X \approx \hat{X} = Y \times_1 U^{(1)} \times_2 U^{(2)} \times_3 \ldots \times_N U^{(N)} \quad (1)$$

The optimal orthogonal matrices $\{U^{(n)}\}_{n=1}^{N}$ and the core tensor $Y$ can be obtained by solving the optimization problem for minimum error between the actual input tensor $X$ and its approximation $\hat{X}$ as:

$$\underset{U^{(1)}, U^{(2)}, \ldots U^{(N)}}{\operatorname{argmin}} \left\| X - Y \times_1 U^{(1)} \times_2 U^{(2)} \ldots \times_N U^{(N)} \right\|_F^2 \quad (2)$$

To initialize the set of basis factor matrices, first, higher order singular vector decomposition (HOSVD) [26] is used, and then, higher order orthogonal iteration is configured to update the orthogonal matrices $\left\{\{U^{(n)}\}_{n=1}^{N}\right\}_{s=1}^{S}$ iteratively until convergence is achieved. Further, the core tensor $Y$ can be obtained through the orthogonal matrices $\{U^{(n)}\}_{n=1}^{N}$ by substituting them into Eq. 2. More explicitly, for our approach, the order of the input tensor is $N = 3$, as shown in Fig. 3, which represents the block from a single knee MRI scan and results in $X \in \mathbb{R}^{I_1 \times I_2 \times I_3}$, with the targeted rank of the output tensor $\{R_1 \times R_2 \times R_3\}$. After computing the initial left singular subspace $U_0^{(n)} \in \mathbb{R}^{I_n \times R_n}$ for $n = \{1,2,3\}$, the optimal solution of $U^{(n)}$ can be obtained as:

$$\underset{U^{(n)}}{\operatorname{argmax}} \left\| U^{(n)T} \times_{-n} \{U^{(n)}\}_{n=1}^{3} \right\|_F^2 \text{ s.t. } U^{(n)T} U^{(n)} = I. \quad (3)$$

By iteratively updating $\left\{\{U_s^{(n)}\}_{n=1}^{3}\right\}_{s=1}^{S}$, the final set of orthogonal matrices can be obtained and the corresponding core

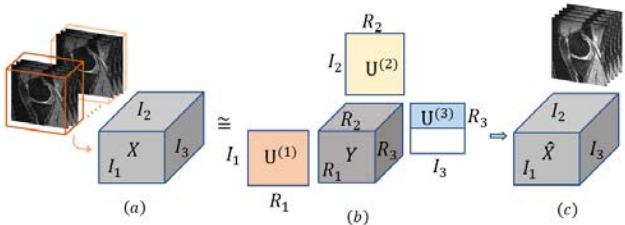

Figure 3: Flowchart of the block-wise tensor construction from knee MRI (a), Tucker decomposition representing the mode matrices and core tensor (b) and resulting reconstruction MRI (c).

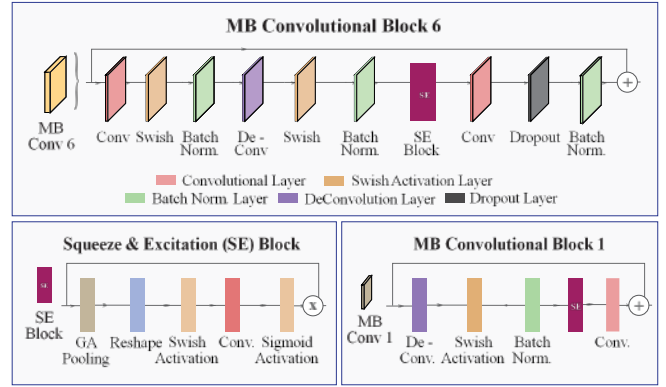

Figure 5: Architecture diagrams of the individual components used in our implementation of encoder decode based segmentation.

tensor $Y \in \mathbb{R}^{R_1 \times R_2 \times R_3}$ can be easily solved by using Eq. 1.

*Low Rank Reconstruction*

By adjusting $R_3$ from the targeted rank $\{R_1 \times R_2 \times R_3\}$ in the decomposed components of the input tensor, structural compression can be easily adjusted while keeping the spatial representation of images unchanged. Finally, the approximation for the original block can be represented in terms of the core tensor $Y$ and the orthogonal matrices $\{U_s^{(n)}\}_{n=1}^{3}$ as:

$$\hat{X} \approx Y \times_1 U^{(1)} \times_2 U^{(2)} \times_3 U^{(3)} \quad (4)$$

### C. Network Architecture

CNNs outperform conventional methods in medical image segmentation, as evident from evaluations on public benchmark datasets [28][42], and they are also reproducible and an order of magnitude faster than traditional multi-atlas and graph-cut segmentation approaches [29]. In particular, U-Nets [11] have shown robust performance in MRI segmentation tasks such as cardiac MRI [30] and in musculoskeletal tissue segmentation [12][14].

1) Segmentation Network

In this paper, we construct a segmentation scheme based on a U-Net architecture. The segmentation network constitutes of symmetrical encoder-decoder architecture instead leveraging compound scaling for contextual information utilizing efficient net [27]. The context information is aggregated using the depth-wise connections while computing complex hierarchical information from the MRI labelled slices. The decoder section of the architecture relies heavily upon the computed information maps and reconstructs the labelled image in a coarse to fine manner.

Fig 4. represents the overview of our encoder-decoder architecture. The design of architecture considers compound scaling method [27] to achieve maximum gains in terms of segmentation outputs. The main building block of the network is MobileNet [40] Convolutional (*MBConv*) block, which efficiently explores squeeze-and-excitation (SE) optimization [39]. These blocks are efficient in terms of semantic information, as they create a shortcut connection between the initial and final layer of convolutional blocks. The activation maps produced at the beginning are expanded to enhance the depth of features while depth-wise convolution [41] is used to lower the channels of final feature map. The increased depth and stacked convolutional blocks aid the learning process by computing more efficient and robust features.

2) Sub-block configuration

In our implementation, the filter-size as well as the number of used blocks are specified on top of each block in diagram, as various convolutional blocks are used in multiple configurations to produce coarse feature maps. The detailed architecture of the *MBConv* blocks and SE block are presented in Fig 5. The *MBConv block-6* block efficiently utilized two convolutional layers at the beginning and one at the end, for exploring the features in SE block, whereas *MBConv block-1* make use of the de-convolutional layer at the beginning, whereas the convolutional layer at the end of SE block. The *MBConv* block constitutes an additional block of swish activation function, which is a combination of linear and sigmoid activations and helps layer to nullify the negative values and improves predictive performance. The exploration of expansion, depth-wise separable convolutions, squeeze and excitation block helps produce the feature maps at bottleneck stage. The *T-MBConv* blocks corresponds to original *MBConv* block however they are reverse in nature and acts as up-sampling layers to form the segmentation maps.

A *softmax* activation function is attached to the last decoder layer to generate the probability map for each tissue class. In our approach, both paths utilize the compound scaling U-Net architecture to train the MRI slices and the corresponding reconstructed MRI slices.

3) Loss function

Considering the class imbalance for the cartilage/meniscus volume when compared with the entire MR volume, our approach uses the weighted cross-entropy loss function between the true segmentation value and output for each model, given by:

$$WCE = \frac{1}{N(X)} \sum_{x \in X} [(t(x) \log(p(x)) w(t(x))]$$ (5)

where $w(t(x))$ is the predefined weight that is computed through class presence from the true value $t(x)$, and $p(x)$ is the predicted output value from the *softmax* function.

D. Alpha Matting and Segmentation

Image matting provides an intuitive way to combine the coarse segmentation predictions and generate a precise segmentation map. In our approach, the output of the multipath DL network is set to generate the automatic *trimap*, which is an estimate of the segmentation, thereby marking the unknown regions. The derivation of a suitable trimap for image matting is complex and difficult; therefore, its direct application in MRI continues to be limited, as careful and low-level exploitation of the boundary clues is critical to adopt from slim tissue regions. In our approach, we address this issue by deriving the automatic trimap directly from DL predictions on the low rank components.

Image matting refers to the task of estimating the foreground object precisely along with the opacity mask for object-level image composition and is a prerequisite in many image editing applications [34]. To perform segmentation on a given image, matting produces a "matte" image that separates the foreground from the background, for which the convex combination can be illustrated through the image composition equation as:

$$I_i = \alpha_i F_i + (1 - \alpha_i) B_i, \alpha_i \in [0,1]$$ (6)

where $F_i$, $B_i$ and $\alpha_i$ represent the foreground, background and alpha matte estimation, respectively, at pixel $i$. For a given input image $I$, the matting process solves $F$, $B$ and $\alpha$ simultaneously; however, the problem is highly ill-posed because the number of unknowns is higher than the number of constraints. This long-studied problem is being addressed by adding more information in terms of additional clues; therefore, matting methods rely on the essential additional input that constrains the solution in terms of a *trimap*. Usually, the trimap is generated manually from user sketches or automatically through the binary segmentation output. The automatically generated *trimap* follows the dilation of the binary mask boundaries; however, in both cases (manual and automatic), the generated *trimap* is coarse. The unknown region in a *trimap* comprises pixels from both opaque and transparent regions because the use of a trimap is tedious for interactive pixel labeling and inaccurate for segmentation methods that utilize dilated binary masks to mark transition regions.

1) Trimap Generation

Precise segmentation of knee tissues and estimation of alpha matte directly from the coarse *trimap* may yield inaccurate tissue boundaries, as targeted regions in knee tissues are in contact with or slightly separated from neighboring tissues and, therefore, may result in inaccurate segmentation. In our proposed pipeline, the segmentation outcome of both the networks is superimposed to generate *trimaps*, defined as:

$$T_i = \begin{cases} F_i \to P_s \wedge P_T \\ u_i \to otherwise \\ B_i \to 1 - [P_s \vee P_T] \end{cases}$$ (7)

where $P_s$ and $P_T$ are the outputs of the segmentation networks from the actual images and the TD images, respectively. The high confidence foreground tissue $F_i$ is derived from the overlap region of both predictions $P_s \wedge P_T$, whereas the High confidence background $B_i$ (non-tissue) is derived from the inclusion of negative class predictions. The unknown or blended region $u_i$ belongs to the remaining pixels and can be further addressed by solving the alpha matting problem.

## 2) Alpha Matte Estimation

From the segmentation perspective, the desired behavior of the alpha matte is to generate the exact 1s and 0s for $F_i$ and $B_i$, respectively, while producing the precise fractional opacity map for the unknown region that ranges between 0 and 1. Matting methods primarily rely on the intensity values of pixels, as well as the position and low-level features, to determine the alpha matte. Several matting methods estimate the alpha matte on the basis of the procedures defined to process the *trimap* information [31][32][33]. Inspired by the manifold structure handling of the learning-based methods, we adopted the alpha estimation described in [33] and extended our trimap structure to estimate the alpha matte. The alpha estimation process determines the value of an individual pixel by assuming a linear combination of its neighborhood pixels, whereas the coefficients of the linear combination can be computed through a local learning process.

For a given pixel location $i$ and its surrounding patch $\mathcal{N}_k$, defined by a window of $3 \times 3$ selected pixels, the alpha $\alpha_i$ at location $i$ can be predicted as a linear function of features within the neighborhood of $\mathcal{N}_k$.

$$\alpha_i = a_k I_i \quad , i \in \mathcal{N}_k \quad (8)$$

where $I_i$ is the intensity feature vector of linear combination coefficients for $i$. Considering the alpha model, the matting problem can be formulated as the constrained optimization problem expressed as:

$$\operatorname*{argmin}_{\alpha_I, a_I} \sum_{k \in I} \sum_{k \in \mathcal{N}_k} (\alpha_i - \sum_{k \in \{I_i\}} a_k I_i)^2 \quad (9)$$
$$\text{subject to} \quad \alpha_k = 1, k \in \Omega_F,$$
$$\alpha_k = 0, k \in \Omega_B.$$

where $\alpha_I$ is the group of all $\alpha$ in image $I$ and $a_I$ is the stacking of linear coefficients $a_k$ contributed by all of the local patches in the image. $\Omega_F$ and $\Omega_B$ represent the foreground and background pixels, respectively, of the trimap regions as an input computed earlier. For every path $k$, the linear coefficients are solved by minimizing the above objective function in (9), whereas the detailed closed-form solution of $a_k$ is addressed in [31]. After substituting $a_k$, the newly derived cost function of $a_I$ can be given by:

$$\mathcal{J}(a_I) = a_I^T L a_I \quad (10)$$

where $a_I$ is reshaped into a vector of length $N$ ($N$ being the total number of pixels in $I$) and $L$ is a matting Laplacian matrix of size $N \times N$. Each $ij^{\text{th}}$ entry corresponding to two different locations $i$ and $j$ in the Laplacian matrix can be obtained from Eq. 11, in terms of the affinity matrix $W_{ij}$.

$$L_{ij} = \begin{cases} W_{ij} & : if\ i = j \\ -W_{ij} & : otherwise \end{cases} \quad (11)$$

The choice of *matting laplacian* varies among different models [33][34][35] and is based on different assumptions. For each pixel $k$ and its neighborhood $\mathcal{N}_k$, a learning-based method [34] is adopted to obtain $W_{ij}$, where $\Sigma_k$ and $u_k$ are the feature covariance matrix and the mean feature vector, respectively, in $\mathcal{N}_k$.

$$W_{ij} = \sum_{\{k: i \in \mathcal{N}_k, j \in \mathcal{N}_k\}} \frac{1}{|\mathcal{N}_k|} [1 + (I_i - u_k)^T \Sigma_k^{-1} (I_j - u_k)] \quad (12)$$

The affinity between pixels $i$ and $j$ is represented by the weight matrix $W_{ij}$, which is characterized by the summation property $W_{ii} = \sum_j W_{ij}$, for $j \neq i$, such that $L$ has zero row summation; therefore, for a specific pair of pixels, the affinity value is higher if they are similar inside the neighborhood in $I$. The higher value eventually helps the labeled alpha values to propagate toward the neighboring similar pixels in unlabeled pixels.

The constrained problem explained above can now be rewritten in the Langrangian form as:

$$\mathcal{J}(\alpha) = \alpha^T L \alpha + \lambda(\alpha - b_S)^T D_S(\alpha - b_S) \quad (13)$$

where $D_S$ is a diagonal matrix of size $N \times N$ and diagonal entries represents the user-constrained pixels; $b_S$ is a vector of length $N$ that comprises the alpha values from the trimap. The closed-form solution is finally solved through a linear system as:

$$(L + \lambda D_S)\alpha = \lambda b_S \quad (14)$$

After this step, alpha matte computes the correspondence of each pixel in the unknown region in terms of the probability map and is further used to distinguish the knee tissues from their neighborhood regions.

## III. EXPERIMENTS AND RESULTS

In this section, we present the experimental results of the proposed method, an analysis and a comparison with other existing methods.

### A. Evaluation Metrics

To evaluate the performance of the proposed method, we used the popular Dice similarity coefficient (DSC), which determines the voxel-wise segmentation similarity between the predicted and the reference ground truth volume. Precision and recall measures were also computed to address segmentation discrepancies. Additional commonly used volumetric metrics such as volume difference (VD) and volumetric overlap error (VOE) were also calculated to indicate size differences.

### B. Datasets

We used datasets from the Osteoarthritis Initiative (OAI). The datasets used in our experiment consisted of 176 sagittal 3D double-echo steady state (DESS) MR datasets, which were collected from 88 subjects, each of whom underwent 2 examinations conducted 12 months apart. Imorphics [21] supplied the manual segmentation of cartilage and meniscus for all these datasets, which were used as the ground truth in our study. The field of view of DESS acquisition was 14 cm with a matrix size of 384×307 (zero-padded to 384×384). We collected 160 slices with slice thickness 0.7 mm. The segmentation problem for cartilage and meniscus was formulated for six subcategorized tissue regions named femoral cartilage (FC), lateral meniscus, lateral tibial cartilage (LTC), medial meniscus (MM), medial tibial cartilage (MTC) and patellar cartilage (PC) on the basis of their occurrence in slices, and this order has been retained in the remaining description of our results.

TABLE I
SEGMENTATION RESULTS FROM INDIVIDUAL KNEE TISSUE COMPARTMENTS

| Metric | Cartilage | | | | Meniscus | |
|---|---|---|---|---|---|---|
| | FC | TC | | PC | M | |
| | | lateral | medial | | lateral | medial |
| Dice | 0.910 (0.018) | 0.919 (0.019) | 0.894 (0.036) | 0.875 (0.044) | 0.884 (0.046) | 0.871 (0.030) |
| Precision | 0.906 (0.020) | 0.932 (0.022) | 0.869 (0.033) | 0.856 (0.039) | 0.911 (0.043) | 0.888 (0.021) |
| Recall | 0.928 (0.025) | 0.938 (0.019) | 0.896 (0.026) | 0.893 (0.038) | 0.911 (0.045) | 0.854 (0.033) |
| VOE | 18.31 (±2.87) | 17.27 (±4.65) | 18.45 (±3.36) | 23.44 (±11.25) | 20.92 (±6.05) | 21.20 (±7.28) |
| VD | 2.547 (±5.01) | 0.339 (±8.40) | 0.825 (±7.11) | -3.376 (±11.49) | 4.140 (±9.04) | -3.777 (±21.39) |

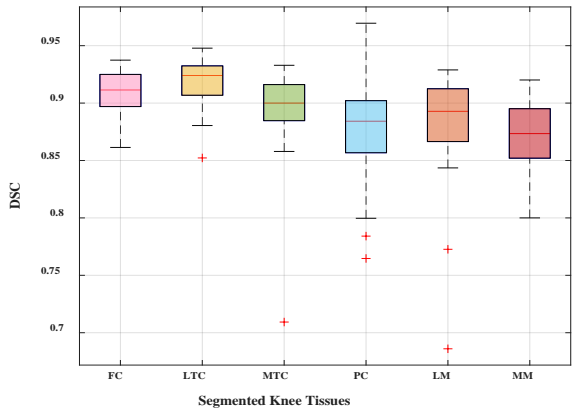

Figure 6: Box plot of the Dice coefficient for each tissue. The median is represented by the line marking the center, and the edges of the box represent the 25[th] and 75[th] percentiles. The maximum and minimum DSCs are represented by the upper and lower extended whiskers, and the outliers are shown in red.

### C. Training and Testing

The proposed method was implemented using Keras v2.2.0 [37] and the Tensorflow v2.2.0 [38] framework on an NVIDIA GTX 1080Ti GPU. All models were trained for 40 epochs by using the Adam optimizer [15] with batch normalization, a learning rate of $10^{-3}$ and a batch size of 8. To avoid overfitting, we adopted standard data-augmentation techniques (axial flips, affine transformations and random crops) in training. All of our methods presented here use training with a 2-fold cross validation scheme. The splitting of a dataset was carried out using the subject IDs from the first and the second half of data and balancing with respect to the severity of the OA status of the subjects, which is provided as their KL grade in the OAI dataset.

### D. Evaluation Results

The evaluation of knee tissue segmentation for 6 structures was carried out using the proposed method, and the average scores (standard deviation) are summarized in Table 1. For the cartilage compartment, FC achieved 0.910, combined TC achieved 0.906 and PC achieved the lowest, 0.875; similarly, VOE and VD for PC were also lower than the corresponding values for other tissues, with higher standard deviation. Figure 6 illustrates the results by showing the box plot for each structure. LTC and FC have the best segmentation score and a low standard deviation; for PC, the standard deviation is high and the cartilage structure has the lowest median. A few outliers with lowest DSC correspond to the subjects with severe cartilage loss due to their late stage of OA. The overall DSC was higher than 0.85 for a majority of the cases in all tissue compartments and was at least 0.8 for all test cases.

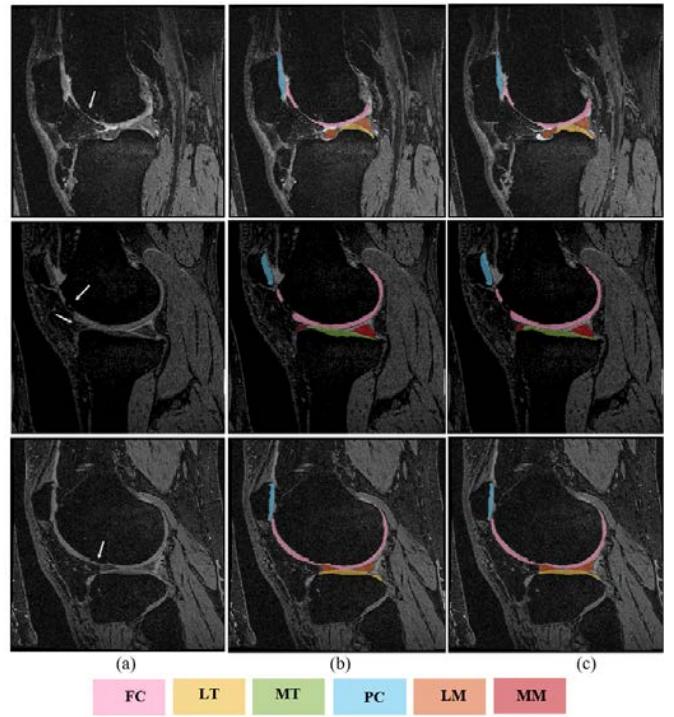

Figure 7: Segmentation results from different subjects, showing the variations in cartilage and meniscus tissues and the corresponding results after using the proposed method.

Segmentation results from the representative slices are presented in Fig. 7. Key regions where the cartilage has thinner connectivity or has a loss of region are marked using arrows. The segmentation results from the proposed method (Fig. 7(c)) show an encouraging adaptation of targeted tissue boundaries in comparison with ground truths (Fig. 7(b)).

To assess the impact of the choice of parameters and matting on the proposed method, we conducted ablation studies with segmentation network using our single path (base) and full pipeline segmentation with trimap generation and matting (combined) for various block sizes and ranks of the decomposed tensor in the proposed method. The results for DSC are presented in Table 2. The proposed method shows improved performance. The optimum Dice score of 0.8925 was achieved for a block size of 10 and a rank value of 3. The cartilage compartment, specifically femoral and tibial (FC, TC), shows a noticeable improvement in the Dice score, as the cartilage is the most affected structure in knee OA and the corresponding changes in cartilage are depicted well in the subsequent frames using *trimap* generation. The block sizes 5, 10 and 15 (out of 160) were tested for rank values 2, 3, 4 and 5. For each block, the best figures have been highlighted in bold. A greater block size has an impact on capturing larger variations, whereas redundancy in information can be controlled by changing the rank value. Different tissue structures have different continuities in the slice dimensions; therefore, they have different structural representations and exhibit corresponding changes for different block sizes. FC has the largest presence in the MRI volume, and for a typical subject in OAI, it constitutes 110 out of 160 slices. Therefore, variations with respect to only the block size are not significant; however, the choice of a suitable rank in each case is vital to generate the precise *trimap*.

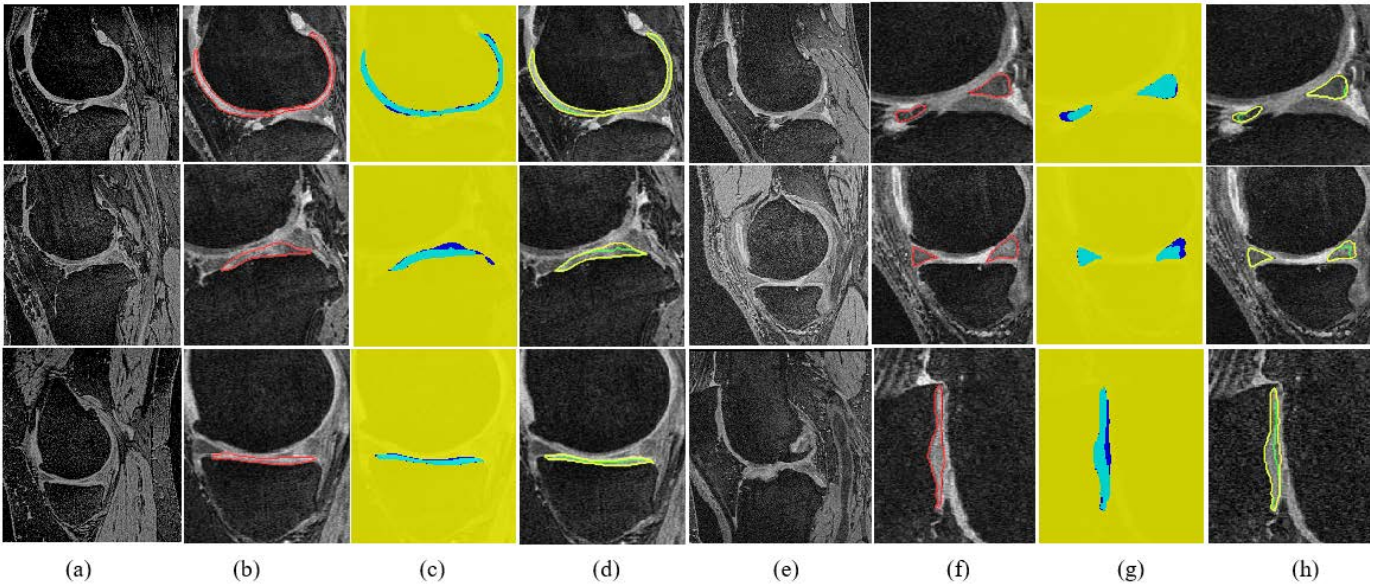

Figure 8: Segmentation results corresponding to each tissue in the cropped regions, indicating the ground truth labels in (b) and (f); proposed *trimap* regions overlaid on image in (c) and (g); and segmentation outcome from the proposed method in (d) and (h). The discrepancies in the segmentation network (in green) are overcome by precise segmentation of the multipath *trimap* network (in yellow).

### E. Trimap Visualization

We propose the generation of a *trimap* directly from the individual network output, which is independent of user intervention and local morphological operations. The captured details are drawn from the temporal correlation among slices and represent only the sparse regions of tissues that undergo changes. The advantage of the method is that it delineates the periphery where the segmentation network overlooks the precise boundary regions. Figure 8 highlights these regions for each tissue and shows the generated *trimap* from multipath segmentation network along with final segmentation outcome. Figures 8(a) and (e) show the input images; Figs. 8(b) and (f) show the cropped tissue regions with overlaid ground truths (shown in red); and Figs. 8(c) and (g) show their corresponding *trimaps* indicating three regions: (1) certain tissue region (cyan), (2) certain non-tissue region (yellow) and (3) unknown region (delineated in blue). The final segmentation results after computing the alpha map (yellow) as well as those with initial segmentation (green) are shown in Figs. 8(d) and (h). It can be observed that tissue boundaries are major contributors to segmentation errors owing to poor tissue contrast; however, through *trimap* generation from multiple paths, it can be precisely recovered and further addressed through alpha matting. The difference between the yellow and green regions, in (c) and (g), indicates the regions retrieved through alpha matting as a final segmentation outcome.

TABLE 2
DICE SCORES COMPARISON WITH DIFFERENT VALUES OF BLOCK SIZE AND RANK

| Path | Block | Rank | Cartilage | | | | Meniscus | | Av. |
|---|---|---|---|---|---|---|---|---|---|
| | | | FC | LTC | MTC | PC | LM | MM | |
| Base | - | - | 0.8844 | 0.8739 | 0.8518 | 0.8540 | 0.8529 | 0.8453 | 0.8604 |
| Combined | 5 | 2 | 0.8901 | 0.8732 | 0.8612 | 0.8588 | 0.8444 | 0.8318 | 0.8599 |
| | | 3 | 0.9030 | 0.8821 | 0.8676 | 0.8541 | 0.8592 | 0.8412 | 0.8679 |
| | 10 | **3** | **0.9104** | **0.9190** | **0.8944** | **0.8753** | 0.8848 | 0.8712 | **0.8925** |
| | | 4 | 0.9087 | 0.9186 | 0.8903 | 0.8642 | **0.8878** | 0.8702 | 0.8900 |
| | | 5 | 0.8913 | 0.9043 | 0.8832 | 0.8536 | 0.8680 | 0.8665 | 0.8778 |
| | 15 | 3 | 0.8717 | 0.8943 | 0.8751 | 0.8638 | 0.8744 | **0.8765** | 0.8760 |
| | | 4 | 0.8891 | 0.9021 | 0.8936 | 0.8621 | 0.8676 | 0.8654 | 0.8800 |
| | | 5 | 0.8834 | 0.8960 | 0.8812 | 0.8656 | 0.8645 | 0.8568 | 0.8746 |

### F. Comparison

We compared our segmentation approach with published state-of-the-art methods in knee tissue segmentation (see Table 3). In terms of DSC, the proposed method performed more accurately than other methods or, for some tissues, as well as other methods. The proposed method with an average DSC of 0.8925 is highest against [12] that used U-Net with DSC of 0.7920 and against [23] with average DSC of 0.8811. In comparison, our method showed an obvious improvement in cartilage segmentation. The method in [23] uses conditional generative adversarial networks (CGANs) in conjunction with U-Net, which results in better segmentation of lateral meniscus. However, the Dice score for this method was lower than that for the proposed method in the case of cartilage. It can also be observed that the proposed method achieved consistent improvement in all knee compartments as compared to other approaches. The main reason behind is the effective use of the secondary input that learns additional clues in thin regions of tissues, whereas other methods emphasized on incorporating multiple CNNs in conjunction with one another.

TABLE 3
DICE SCORES COMPARISON WITH STATE-OF-THE-ART METHODS

| Methods | Cartilage | | | | Meniscus | |
|---|---|---|---|---|---|---|
| | FC | TC lateral | TC medial | PC | M lateral | M medial |
| Norman et al. [12] | 0.867 (0.032) | 0.799 (0.036) | 0.777 (0.029) | 0.767 (0.091) | 0.812 (0.030) | 0.731 (0.054) |
| Ambellan et al. [17] | 0.894 (0.024) | 0.904 (0.024) | 0.861 (0.053) | - | - | - |
| Egor et al. [36] | 0.907 (0.019) | 0.897 (0.028) | | 0.871 (0.046) | 0.863 (0.034) | |
| Sibaji et al. [23] | 0.8967 (0.023) | 0.9177 (0.013) | 0.8608 (0.039) | 0.8426 (0.061) | **0.8919** (0.024) | 0.8708 (0.045) |
| **Proposed** | **0.9104** (0.018) | **0.9190** (0.019) | **0.8944** (0.036) | **0.8753** (0.044) | 0.8848 (0.046) | **0.8712** (0.030) |

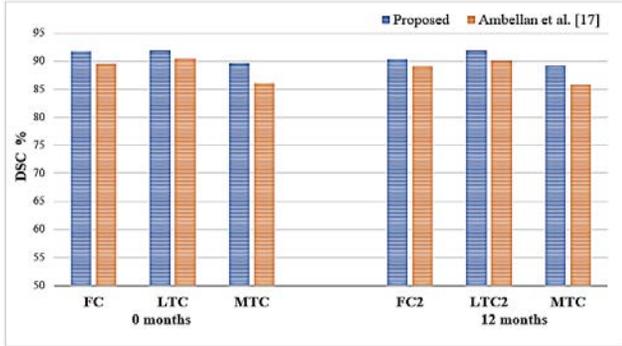

Figure 9: 12-month follow-up segmentation performance comparison with other method.

Another implementation of knee tissue segmentation that uses a combination of 2D and 3D CNNs along with SSMs [17] was included in the comparison to highlight the applications in longitudinal studies. Subjects with an initial scan and a 12-month follow-up were examined. The segmentation accuracy was assessed based on the changes that occur in cartilages over time. The results are presented in Fig. 9. The proposed method achieved consistent improvement in segmentation of cartilage structure when compared with the state-of-the-art method. This result shows that the progression and severity of OA may have little effect on segmentation, which is desirable for the analysis and quantification of OA in clinical practice.

## IV. APPLICATION IN CARTILAGE THICKNESS

Articular cartilage degeneration and loss are considered to be critical elements critical elements in the pathophysiology of OA. To understand the impact of disease on articular cartilage, it is beneficial to understand the discrepancies in physiological cartilage thickness. Therefore, an accurate measurement of cartilage thickness will be advantageous for both detection and monitoring purposes. As an application, we also investigated the segmentation performance on a separate MRI sequence (T1-FFE or SPGR). We expected our segmentation to be more challenging for a new, unseen subject with a different MRI sequence. The qualitative segmentation results for the subject are presented in Fig. 10; the effectiveness of the proposed method in segmentation can be noted. We investigated the segmentation outcome of the proposed method to render the thickness map of the articular cartilage. Fig. 10(a) (last row) show the 3D rendering of the cartilage tissue and the corresponding thickness map of the tissue Figs. 10(b-c). The cartilage regions at the edges as well as those with thinner regions or missing regions can be easily observed using the thickness map.

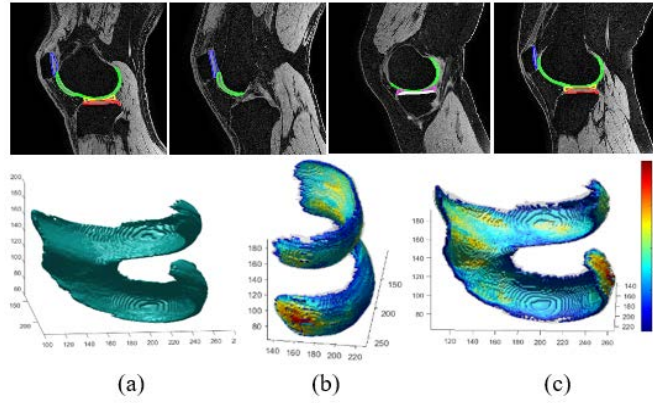

Figure 10: An illustration of the application of proposed segmentation method on knee tissue segmentation and thickness map from OA subject. Qualitative segmentation (top rows), (a) 3D rendering of cartilage, (b) and (c) cartilage thickness map for different view angle.

## V. CONCLUSION

We propose an automatic segmentation framework for knee tissue segmentation that utilizes multipath CNN networks with distinct inputs and fuses them by using an image matting–based label refinement mechanism. Low rank Tucker decomposition is performed for each sub-block to train the secondary path for CNNs. The key features of the proposed method are the construction of the low rank representation of local blocks to capture the intrinsic structural similarity and CNN prediction based on secondary path to generate the intermediate confidence region. The proposed trimap generation addresses the segmentation problem by using image matting. From a visualization of the *trimaps* for knee tissues, it is observed that shape consistency in the 3D format for knee tissues and boundary-related discrepancies for different patients are effectively modeled through low rank tensor decomposition, which is well suited to describe interdimensional structures.

The method was validated by using a publicly available OA dataset. The segmentation results from the proposed method demonstrated reasonable agreement with the tissue contours for ground truth. Moreover, the Dice-score performance for subjects with different KL grades indicated the robustness of the proposed method for the OA severity scale. Considering the efficacy of the proposed method in terms of obtaining segmentation masks in knee tissues and generating a valid thickness map, we would like to extend our work for the quantitative analysis of OA.